# An Overview of the Effect of Self-Gravity on the Structure of Accretion Discs

D. Saikia[1], L. Devi[2,*], B. Sarkar[2], A. J. Boruah[2]

[1]Department of Mechanical Engineering, Tezpur University, Tezpur - 784028, Assam, India

[2]Department of Applied Sciences, Tezpur University, Tezpur - 784028, Assam, India

*Corresponding author, Email: app23110@tezu.ac.in

**Abstract.**

In astrophysical systems like X-ray binaries (XRBs), active galactic nuclei (AGN), and young stellar objects (YSOs), we often observe a very fundamental structure called accretion discs (ADs). Conventional AD theory usually supposes that the gravitational field is controlled by a central compact object. This assumption breaks down when the mass of the disc becomes considerable in contrast to that of the massive central object. In these cases, the AD's self-gravity (SG) can drastically change its structure, dynamics, and evolution. This review investigates how SG influences the radial and vertical structure of ADs and how it modifies the mechanisms that transport angular momentum (AM). Along with these, this review also tries to explore how gravitational instabilities (GIs) evolve and how they affect disc fragmentation and astrophysical phenomena like stellar and planetary formation, AGN dynamics, and gamma-ray bursts (GRBs).

## 1 Introduction

ADs are gaseous, differentially rotating formations that carry matter inward toward a central object [1, 2]. In doing so, AM must be redistributed outward. Classical models, such as the thin disc model developed by Shakura & Sunyaev [3], assume optically thick, geometrically thin discs dominated by the gravitational potential of a massive central object. These assumptions work well in the inner regions of black hole (BH) ADs but fail in systems with discs with high mass or larger radius. In such regimes, the disc's own gravity becomes non-negligible, such as in protostellar environments or galactic nuclei.

Disc SG becomes dynamically important when the disc's own gravitational potential begins to influence the gas motion, either in the vertical direction (affecting the vertical hydrostatic balance and scale height) or in the radial direction (modifying the rotation curve[†] and disc stability) [4]. It is essential to understand the AD's global structure using SG for comprehending various astrophysical phenomena, including star and planet formation, supermassive BH (SMBH) feeding, and episodic outbursts in AGN and GRBs [4, 5].

AD theory has primarily focused on non-self-gravitating cases, without considering the SG effects. In the last ten years, recognition of the significance of disc SG has grown significantly [6, 7, 8]. Observational evidence increasingly suggests that disc masses can be comparable to or even exceed the central object's mass fraction in YSOs and galactic nuclei, making SG a key ingredient in shaping disc structure and triggering phenomena such as fragmentation or star formation. At the same time, theoretical and computational improvements enable thorough analysis of GIs and their nonlinear evolution. However, most conventional disc models ignore SG, skipping out essential physics in regimes where it dominates. In addition to providing an explanation for the observed disc variability and structure, filling this gap would help connect small-scale accretion physics to large-scale astrophysical phenomena. These developments strongly motivate a deeper exploration of SG in ADs, which is the focus of this work.

---

[†]A rotation curve is a plot that shows how the orbital velocity of an object (such as a star or a cloud of gas) changes with respect to its distance from the center of a rotating system [9, 10].

## 2    Vertical and Radial Components of SG

Traditionally, vertical SG has been the primary focus in analytical treatments of ADs [8]. Paczyński [8] introduced an expression for the vertical component of SG, showing that it becomes significant at a radius where the parameter $A = \frac{4\pi\rho r^3}{M}$ (where $A$ is dimensionless parameter indicating the importance of vertical SG in an AD; $r$ is the radial distance from the central object; $\rho$ is disc's mass density (at a given r); $M$ is central object's mass) exceeds unity [2]. The free parameter $A$ cannot be determined from the disc structure. The scale height and density distribution of the disc are changed because of the inclusion of vertical SG. Lodato [4] showed that the mass of the AD should be in proportion to the mass of the central object to affect the SG in the radial direction. Using models of self-regulating ADs (a self-regulating disc is one for which efficient cooling processes are supposed to keep the disc near Jeans' stability margin[#] [11]), this can be confirmed [6]. The Keplerian approximation is typically applicable when the AD mass is significantly lower than the central object. Non-Keplerian rotation has been detected in a few significant cases, though they can be explained in terms of comparatively massive discs. To determine the disc's stability, its vertical gravity has a major impact. Lodato [4] also showed that even when the disc's mass is far less than the central object's mass, the disc SG still affects the disc's vertical structure for thin discs.

## 3    Viscosity and AM Transport

The fundamental process that enables accretion in discs is AM transport. In an AD, for matter to spiral inward (accrete), it must lose AM. The AM must be transported outward to other parts of the disc. Viscosity provides the mechanism for this transport. Viscosity in ADs differs from the simple molecular viscosity of typical fluids, which is too tiny to impact ADs. Instead, it refers to effective viscosity. It acts as an internal "friction" between neighboring fluid layers moving at slightly different velocities. This velocity gradient creates shear. This shear stress transfers AM outward while allowing matter to lose AM and drift inward. A widely used model is the $\alpha$-prescription of viscosity by Shakura & Sunyaev [3], where the effective kinematic viscosity is defined as $\nu = \alpha c_s h$. Here, $h$ is the disc's scale height and $c_s$ is the sound speed [1, 13, 3]. The famous Shakura & Sunyaev $\alpha$-viscosity model faces constraints in self-gravitating regimes [3, 14]. The $\alpha$-disc model merely incorporates physical constraints on the turbulence's efficiency in a disc, not the mechanism that creates it. Indeed, it can be anticipated that any cosmic shear flow with a high Reynolds number ($R_e$) would show some form of turbulent viscosity, whether within a disc or not. Thus, it can be said that the $\alpha$-model for ADs is based on a more general prescription [14].

Current theoretical models of ADs do not fully capture the physics behind the viscosity. This issue is essential because thorough modeling of the formation and evolution of ADs relies on the viscosity and its relationship to physical parameters [14].

---

[#]Jean's stability margin describes a limit above which gas cloud becomes gravitationally unstable and collapses, leading to star formation [12].

In the year 2000, Duschl et al. [14] suggested an alternative viscosity prescription. This prescription was termed as *β*–viscosity. In this alternative prescription, the viscosity is parameterized to be proportional to both the radius ($R$) and the azimuthal velocity ($\Omega$). In a Keplerian disc (KD), this relation is equivalent to $\nu_\beta = \beta\Omega R^2$. This prescription was originally proposed by Lynden-Bell & Pringle [15], and some aspects of it have been discussed in later papers [16, 17, 18, 19, 20]. This *β*-prescription was revived by Duschl et al. [21, 14] since the only relevant scales in a KD are the angular velocity and the radius, which contain all the information about the rotation and the curvature of the flow. Assuming that the critical $R_e$ is the same as in plane-parallel shear flows (i.e., $R_e \approx 10^3$ and hence $\beta = 10^{-3}$), the authors make the effective decision to equate the parameter *β* with the inverse of $R_e$ [14]. One notable characteristic of this *β*-viscosity is that it is independent of local physical conditions and depends on the radius of a KD. Conversely, the conventional *α*-prescription depends on the local values of sound speed $c_s$ and scale height $h$ [22]. This prescription produces realistic models of ADs that account for both Keplerian motion and SG. Furthermore, Duschl et al. [14] recover the *α*-disc solution as a limiting case for thin discs with a suitably small mass. The transport of AM in self-gravitating discs is better described by this model than the *α*-viscosity model. The *α*-disc model is considered to be inconsistent when the mass of the disc is large enough for SG to play any role. This inconsistency can be solved by the *β*-disc model. In Keplerian self-gravitating discs, this issue still occurs since the azimuthal motion is still Keplerian, but only the vertical structure is controlled by SG. Moreover, it was shown by Liu et al. [23] that *β*-viscosity more accurately describes the global dynamics of AGN discs, particularly in the outer regions where SG is more important. While SG plays a part in AGN discs, it does so at a considerable distance from the core object [24]. In the outer regions of AGN ADs, the gas is cool and often becomes self-gravitating. Hence, $c_s$ decreases as we move outward from the center. Consequently, very low viscosities are predicted by the *α*-prescription, and AM transport nearly stops. However, the *β*-viscosity model is independent of the thermal state and only depends on the disc's rotation and geometry. At large radii, $r^2$ is enormous, and even though $\Omega$ decreases with $r^{-\frac{3}{2}}$, the product $r^2\Omega \sim r^{\frac{1}{2}}$ increases with radius. This indicates that, regardless of $c_s$, viscous transport in the outer disc naturally becomes stronger.

Kubsch et al. [25] used self-similar models to study α-viscosity in self-gravitating discs. In the context of AGN discs, they examine *α*-viscosity's effectiveness in self-similar models that account for the disc's SG. The change in the overall mass distribution with time of SMBH in AGN cannot be explained by *α*-viscosity; yet it appears appropriate for modeling protoplanetary discs (PPDs).

## 4    Self-Similar and Semi-Analytical Models

The global structure of self-gravitating discs is often studied using self-similar solutions [6, 7, 26, 27]. Mineshige & Umemura [7] developed a steady-state self-similar model with a self-gravitating disc in both vertical and radial directions. Their idea generates a smooth rotation curve and a constant temperature. This solution provides that equilibrium angular velocities ($\Omega$) change as a power-law function of the radius ($\sim r^{-\gamma}$) (where $\gamma$ is the free parameter). These solutions are helpful for developing intuition regarding the scaling of physical quantities in various regimes, for example, in AGN and YSOs, where the conventional thin-disc approximation is ineffective [7]. According to Illenseer et al. [26], the power-law ($\sim r^{-\gamma}$) is the most basic monotone function that might be considered a suitable rotation law. In two instances, the monopole approximation is accurate: $\gamma = -\frac{3}{2}$ and $-1$. The latter value is related to Mestel's disc[#], while the former is associated with Keplerian rotation. With the monopole approximation, the error would be less than 10% for $-\frac{3}{2} < \gamma < -1$. The gravitational acceleration acting inward is overestimated by less than a factor 2 for $\gamma = -\frac{3}{4}$. This finding supports Lin & Pringle's [28] observation that an accuracy of around 5% can be expected with the monopole approximation. If the $\gamma$ gets close to $-\frac{1}{2}$, the monopole approximation fails.

Illenseer et al. [26] also suggested that γ not only characterizes the rotation curve, but through its effect it also governs how the enclosed mass and the surface density distribution behave with radius in the AD's outer region. A rotation law with a greater gradient results in a flatter radial growth of the enclosed mass by causing surface density to fall off more steeply with radius. Therefore, the disc's γ basically serves as another self-gravitating component. The authors thus draw the conclusion that the central object will grow rapidly the more self-gravitating the AD is. Additionally, they have used their model on an AD from an AGN. They also discussed how SMBHs arise in this context and discovered that a rapid rise of SMBHs in the early universe could be explained by their model of self-gravitating ADs. Self-gravitating models are essential to study the dynamics of large ADs where the gravitational influence of the central object is not the only dominant force. They also offer a more realistic framework for investigating physical quantities such as temperature, density, and rotation scale with radius.

Abbassi et al. [29] extended self-similar models to incorporate winds and polytropic structures, and reported that SG dramatically alters the profiles of accretion rate, scale height, and surface density. By considering the disc's self-gravitational field and using the $\alpha$-viscosity and $\beta$-viscosity model in a geometrically thin AD rotating around a compact object sitting at the center, the authors have investigated the effects of hydrodynamical winds. According to their research, a significant portion of the disc experiences GI, as defined by the Toomre $Q$-parameter. $Q < 1$ suggests that the disc is prone to fragmentation [29]. $Q$-parameters were found to increase with wind strength, suggesting the disc is more stable. Nevertheless, there are differences in how the $Q$-parameter depends on the wind parameter. Even when wind is present, the GIs in $\beta$-discs are more noticeable than in $\alpha$-discs. They therefore expected the planet formation surrounding newborn stars to be described more effectively by the $\beta$-model. The spectra produced by $\beta$-prescription for PPDs are far flatter than those produced by non-self-gravitating discs, which is more consistent with observations [30].

Except in the cases where $l = 0, 0.5$ ($l$ is the AM parameter of the wind), the results show that the $Q$-parameter rises with winds or outflows [29]. There are unphysical solutions for $l = 0, 0.5$. Typically, in an $\alpha$-model, $Q > 1$, especially when winds are present, except in the inner portion of the disc. Regarding the viscosity parameterization using the $\beta$-parameter, they found that, in the presence of a typical wind, most disc regions have $Q$ values below the threshold for instability ($Q_{thr} \approx 1$) [29].

## 5  GI and Fragmentation

One characteristic of self-gravitating discs is GI, which has major significance during their evolution. The Toomre parameter, $Q = \frac{c_s k}{\pi G \Sigma}$, is used to quantify the stability developed against axisymmetric perturbations [31]. Here, $k$ and $\Sigma$ are the disc's epicyclic frequency and surface density, respectively, and $G$ is the gravitational constant. The stability condition Q > 1, known as Toomre's criterion, extends the Jeans stability analysis to rotating ADs. Notably, it accounts for the SG of the AD material. The self-gravitational pull of the disc material can be countered by the combined actions of rotating shear (epicyclic frequency) and pressure support (sound speed). In gaseous discs, $Q$ is often represented as $Q_{gas}$; values exceeding 1 suggest stability, implying minimal star formation in molecular clouds. When $Q < 1$, the disc becomes unstable, and fragmentation starts. This instability, as a result, forms the spiral arms, clusters, or even cause the collapse of material into dense cores that might eventually give rise to stars or planetesimals [4]. In massive systems like YSOs or AGN, the traditional idea of a thin, stable disc doesn't quite fit. That's because the disc's gravity becomes increasingly significant, which changes the behavior and evolution of the disc [29]. Fig. 1 shows how the stability parameter $Q$ changes across the disc radius at different times during a simulation of a disc with mass $0.5 M_\odot$. The disc is not yet in steady state. It's going through an evolving, unstable phase where instabilities are growing or decaying. $M_{disc} = 0.5 M_\odot$ i.e., the disc mass is half the central star's mass. Since it's a very massive disc, SG strongly affects stability. All times are expressed in units of the Keplerian rotation period of the outer disc $t_0 = \frac{2\pi}{\Omega_{out}}$. The solid line refers to $t = 1.9 t_0$ (Early stage of the transient), the dotted line refers to $t = 2.7 t_0$ (Middle stage), and the dashed line refers to $t = 7 t_0$ (Later stage of the evolution). Initially $Q$ is low (unstable), then the profile evolves as the transient develops, and the disc finally tends toward a more stable state [32].

---

[#]A disc which extends out to infinity and follows that the radial distance between the object seated in the center and a point in the disc is inversely related to the surface density at that point [4].

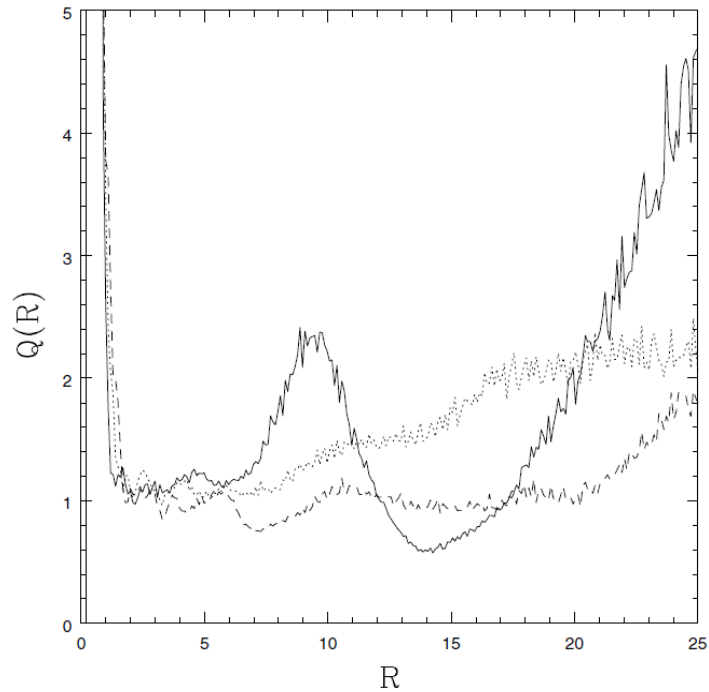

Figure 1: Evolution of the $Q$-parameter across the disc radius at different times. The three curves refer to $t = 1.9 t_0$ (solid line), $t = 2.7 t_0$ (dotted line), and $t = 7 t_0$ (dashed line) [from figure 2 of Lodato & Rice [32]].

The GI and fragmentation occur in the cooling time scale. As shown by Rice et al. [33] & Grammie [34], the condition for fragmentation depends on the AD cooling time ($t_{cool}$) satisfying

$$\Omega t_{cool} < \xi \qquad (1)$$

where $\xi$ is a parameter, whose value is ~1, but from numerical simulation $\xi \approx 3$ [35].

Lodato [4] and Lodato & Rice [32] the studied nonlinear evolution of GIs in discs. The transmission of AM through the disc may be significantly facilitated by GIs. The disc could fragment into gravitationally bound clumps as a result of the instability's nonlinear consequence, which can be rather violent under some circumstances [4]. Those bound clumps eventually become stars or planets if the cooling timescale is short enough. Recent studies on self-gravitating discs have mostly focused on examining the circumstances in which stability develops as a long-lived spiral structure that either determines the disc fragmentation or facilitates the transport of AM. The areas with high density within the disc may eventually break apart and collapse into self-gravitating clumps due to the spiral arms that arise [4, 32]. Hence, the interplay between GI and AM transport is essential to understand the long-term evolution of ADs in various astrophysical systems, including AGN and protostellar discs.

## 6    Observational Implications

It is necessary to understand the impact of SG on ADs while studying the dynamics of YSOs, protostellar and PPDs, GRBs, and AGNs. SG can explain various observable phenomena, including spiral arms in discs, high-energy bursts in GRBs, and variability in AGNs.

According to Goodman [5], standard steady-state ADs surrounding quasi-stellar objects (QSOs) may fragment into stars due to SG, rather than feeding the central BH [5]. Rafikov et al. [35] proposed GI as a probable cause of large planet formation in PPDs. Their analysis suggests that only GI in PPDs is insufficient for direct planet formation. There must be cooling mechanisms by radiative losses from PPD's photosphere, which follows equation (1). Now, in this case, the $t_{cool}$ can also be represented as a measure of thermal energy versus radiative loss, i.e.

$$t_{cool} \approx \frac{\Sigma c_s^2}{\gamma-1} \frac{f(\tau)}{2\sigma T^4}, f(\tau) = \tau + \frac{1}{\tau} \qquad (2)$$

where, T is the temperature of the midplane, σ is the Stephan-Boltzmann constant, $\tau$ is the optical depth of the disc, and $f(\tau)$ characterizes the effectiveness of cooling when $\tau \approx 1$ [35]. The necessary condition for planet formation by GI is

$$(\Sigma \frac{f(\tau)}{\zeta} \frac{\Omega}{\sigma} (\frac{k}{\mu})^4)^{\frac{1}{6}} \leq c_s \leq \pi Q_0 \frac{G\Sigma}{\Omega} \qquad (3)$$

(All the notations have the same meaning as Rafikov et al. [35])

When condition (3) is fulfilled, discs become unstable and sufficiently fast cooling allows for fragmentation following equation (2) to form confined gaseous clumps that eventually disintegrate to become planets [35]. Durisen et al. [36] found that GI in protostellar and PPDs may cause fragmentation and AM transport during rapid cooling. Kratter & Lodato [37] suggest that GI controls the evolution of the early disc and promotes planet formation through migration and cluster formation. Simulations by Mayer et al. [38] showed that for a cooling time scale equal to the orbital timescale, isolated large discs ($M \sim 0.1 M_\odot$) fragment into protoplanets due to GI. PPDs are less likely to fragment in areas closer to the protostar than a critical radius ($r_c$) due to high temperatures and strong Coriolis forces. Their findings suggest that fragmentation around the center of a star is difficult but not impossible. The PPD circling Elias 2-27 (A young 0.8 Myr M0 star, situated in the Ophiucus star-forming region around 115 parsec away) was recently observed using Atacama Large Millimeter/submillimeter Array (ALMA), and the results showed a two-armed spiral configuration. With the stars' young age, observable morphology, and the disc-to-star mass ratio inferred from dust continuum emission, this system is ideal for studying the function of SG in the early stages of star formation [39].

Palit et al. [40] studied the implications of the AD's SG surrounding a BH linked to long GRBs in an evolving background Kerr metric. They contended that the reported GRB characteristics and jet production are influenced by SG, which causes local instabilities and modifies the final

BH spin and mass. The findings of Janiuk et al. [41] demonstrate that changes in the GRBs' prompt emission and flare activity can be explained by GI. The authors found that in the innermost regions of the collapsing star, the GI, as measured by $Q$, has a substantial impact. The brightest GRBs exhibit rapid fluctuation, which agrees with the self-gravitating collapsar concept.

Self-gravitating ADs play a crucial role in powering SMBHs in AGN. Illenseer et al. [26], modeled a self-gravitating AD and applied it to an AGN system. They found that self-gravitating ADs may explain the fast emergence of SMBHs in the universe's early stage (like AGN).

In high-energy astrophysical contexts like compact binary mergers or collapsars, discs become dense and hot enough for neutrino cooling to dominate. Liu et al. [23] explored how SG affects the dynamics and structure of neutrino-dominated accretion flows (NDAFs), especially in the outer region of the disc. They also reviewed that the Toomre $Q$ parameter's vertical distribution can trigger the disc's GI, which may explain GRB's late time flares and extended emission in short-duration GRBs.

# 7    Conclusions

In this review, we conclude that SG profoundly alters the structure and evolution of ADs. In contrast to traditional models, self-gravitating discs exhibit non-Keplerian rotation and complex vertical structure and are susceptible to GIs that drive AM transport and fragmentation. The incorporation of realistic viscosity prescriptions, particularly those grounded in turbulent hydrodynamics, is essential for accurate modeling.

Applications of SG span a broad range of astrophysical phenomena, from star and planet formation in protostellar discs to SMBH growth and episodic activity in AGN, and energy generation in GRBs. Future advancements in observational capabilities, combined with high-resolution simulations, will further elucidate the intricate role of SG in disc physics.

Also, the interplay of SG, radiative processes, and magnetic fields is an active research area in the study of ADs. When gravitational physics and magnetohydrodynamic (MHD) processes are combined, disc dynamics can be better understood, especially in regimes where turbulence, outflows, and jet formations occur. Interpreting multi-wavelength disc observations in many astrophysical contexts requires this multidimensional approach.

There's a need for refinement in theoretical models of ADs having SG to meet the progressively more precise views of disc morphology and kinematics provided by observational methods using ALMA, James Webb Space Telescope (JWST), and the future Laser Interferometer Space Antenna (LISA) and Extremely Large Telescope (ELT) [42, 43].

**Acknowledgement**

The authors DS, LD, BS, and AJB acknowledge the reviewer for thoroughly reviewing the manuscript and providing beneficial comments.

# References

[1] U. Kolb, *Extreme environment astrophysics*. 2010.

[2] L.-T. Yang and X.-C. Liu, "*Radial self-gravity of accretion discs around supermassive black holes.*," Acta Astrophys. Sin., **vol. 10, no. 3**, pp. 199–208, 1990.

[3] N. I. Shakura and R. A. Sunyaev, "*Black holes in binary systems. Observational appearance.*," Astron. Astrophys., **vol. 24**, pp. 337–355, 1973.

[4] G. Lodato, "*Self-gravitating accretion discs*," Riv. Nuovo Cimento, no. Online First, pp. 293–99999, Dec. 2007, doi: 10.1393/ncr/i2007-10022-x.

[5] J. Goodman, "*Self-gravity and quasi-stellar object discs*," Mon. Not. R. Astron. Soc., **vol. 339, no. 4**, pp. 937–948, 2003.

[6] G. Bertin and G. Lodato, "*A class of self-gravitating accretion disks*," Astronomy and Astrophysics, **vol. 350**, pp. 694–704, 1999, Accessed: Jun. 09, 2025. [Online].

[7] S. Mineshige and M. Umemura, "*Self-similar, self-gravitating viscous disks*," Astrophys. J., **vol. 469, no. 1**, p. L49, 1996.

[8] B. Paczynski, "*A model of selfgravitating accretion disk*," Acta Astron., **vol. 28, no. 2**, pp. 91–109, 1978.

[9] G. Lodato and G. Bertin, "*Probing the rotation curve of the outer accretion disk in FU Orionis objects with long-wavelength spectroscopy*," Astron. Astrophys., **vol. 408, no. 3**, pp. 1015–1028, 2003.

[10] T. S. Van Albada and R. Sancisi, "*Dark matter in spiral galaxies*," Philos. Trans. R. Soc. Lond. Ser. Math. Phys. Sci., **vol. 320, no. 1556**, pp. 447–464, 1986.

[11] G. Bertin, "*Self-regulated accretion disks*," Astrophys. J., **vol. 478, no. 2**, p. L71, 1997.

[12] R. S. Klessen, "*Star formation in molecular clouds*," EAS Publ. Ser., **vol. 51**, pp. 133–167, 2011.

[13] J. Frank, A. R. King, and D. Raine, *Accretion power in astrophysics*. Cambridge university press, 2002.

[14] W. J. Duschl, P. A. Strittmatter, and P. L. Biermann, "*A note on hydrodynamic viscosity and selfgravitation in accretion disks*," Astronomy and Astrophysics, **vol. 357**, pp. 1123–1132, 2000.

[15] D. Lynden-Bell and J. E. Pringle, "*The evolution of viscous discs and the origin of the nebular variables*," Mon. Not. R. Astron. Soc., **vol. 168, no. 3**, pp. 603–637, 1974.

[16] J. A. de Freitas Pacheco and J. E. Steiner, "*The spectrum of Cyg X-1—A theoretical model*," Astrophys. Space Sci., **vol. 39, no. 2**, pp. 487–494, 1976.

[17] R. I. Thompson, P. A. Strittmatter, E. F. Erickson, F. C. Witteborn, and D. W. Strecker, "*Observation of preplanetary disks around MWC 349 and LKH-alpha 101*," Astrophys. J., **vol. 218**, pp. 170–180, 1977.

[18] D. N. C. Lin and J. Papaloizou, "*On the structure and evolution of the primordial solar nebula*," Mon. Not. R. Astron. Soc., **vol. 191, no. 1**, pp. 37–48, 1980.

[19] R. E. Williams, "*Emission lines from the accretion disks of cataclysmic variables*," Astrophys. J., **vol. 235**, pp. 939–944, 1980.

[20] I. Hubeny, "*Vertical structure of accretion disks-A simplified analytical model*," Astrophys. J., **vol. 351**, pp. 632–641, 1990.

[21] W. J. Duschl, P. A. Strittmatter, and P. L. Biermann, "*Hydrodynamic Viscosity and Self-gravitation in Accretion Disks*," in American Astronomical Society Meeting Abstracts# 192, pp. 66–17, 1998.

[22] J.-M. Huré, D. Richard, and J.-P. Zahn, "*Accretion discs models with the β-viscosity prescription derived from laboratory experiments*," Astron. Astrophys., **vol. 367, no. 3**, pp. 1087–1094, 2001.

[23] T. Liu, X.-F. Yu, W.-M. Gu, and J.-F. Lu, "*Self-gravity in neutrino-dominated accretion disks*," Astrophys. J., **vol. 791, no. 1**, p. 69, 2014.

[24] I. Shlosman and M. C. Begelman, "*Self-gravitating accretion disks in active galactic nuclei*," Nature, **vol. 329, no. 6142**, pp. 810–812, 1987.

[25] M. Kubsch, T. F. Illenseer, and W. J. Duschl, "*Accretion disk dynamics-α-viscosity in self-similar self-gravitating models*," Astron. Astrophys., **vol. 588**, p. A22, 2016.

[26] T. F. Illenseer and W. J. Duschl, "*Self-similar evolution of self-gravitating viscous accretion discs*," Mon. Not. R. Astron. Soc., **vol. 450, no. 1**, pp. 691–713, 2015.

[27] M. Shadmehri and F. Khajenabi, "*A class of self-gravitating, magnetized accretion disks*," Astrophys. J., **vol. 637, no. 1**, p. 439, 2006.

[28] D. N. Lin and J. E. Pringle, "*The formation and initial evolution of protostellar disks*," Astrophys. J., **vol. 358**, pp. 515–524, 1990.

[29] S. Abbassi, E. Nourbakhsh, and M. Shadmehri, "*Viscous accretion of a polytropic self-gravitating disk in the presence of wind*," Astrophys. J., **vol. 765, no. 2**, p. 96, 2013.

[30] S. Abbassi and J. Ghanbari, "*Beta Viscose Prescription in Self-Gravitating Disks*,", 2009.

[31] A. Toomre, "*On the gravitational stability of a disk of stars*," Astrophys. J., **vol. 139**, pp. 1217–1238, 1964.

[32] G. Lodato and W. K. M. Rice, "*Testing the locality of transport in self-gravitating accretion discs—II. The massive disc case*," Mon. Not. R. Astron. Soc., **vol. 358, no. 4**, pp. 1489–1500, 2005.

[33] W. K. M. Rice, P. J. Armitage, M. R. Bate, and I. A. Bonnell, "*The effect of cooling on the global stability of self-gravitating protoplanetary discs*," Mon. Not. R. Astron. Soc., **vol. 339, no. 4**, pp. 1025–1030, 2003.

[34] C. F. Gammie, "*Nonlinear Outcome of Gravitational Instability in Cooling, GaseousDisks*," Astrophys. J., **vol. 553, no. 1**, p. 174, 2001.

[35] R. R. Rafikov, "*Can giant planets form by direct gravitational instability?*," Astrophys. J., **vol. 621, no. 1**, p. L69, 2005.

[36] R. H. Durisen, A. P. Boss, L. Mayer, A. F. Nelson, T. Quinn, and W. K. M. Rice, "*Gravitational instabilities in gaseous protoplanetary disks and implications for giant planet formation*," Protostars Planets V, p. 607, 2007.

[37] K. Kratter and G. Lodato, "*Gravitational instabilities in circumstellar disks*," Annu. Rev. Astron. Astrophys., **vol. 54, no. 1**, pp. 271–311, 2016.

[38] L. Mayer, J. Wadsley, T. Quinn, and J. Stadel, "*Gravitational instability in binary protoplanetary discs: new constraints on giant planet formation*," Mon. Not. R. Astron. Soc., **vol. 363, no. 2**, pp. 641–648, 2005.

[39] B. Veronesi *et al.*, "*A dynamical measurement of the disk mass in Elias 2–27*," Astrophys. J. Lett., **vol. 914, no. 2**, p. L27, 2021.

[40]  I. Palit, A. Janiuk, and P. Sukova, "*Effects of self-gravity of the accretion disk around rapidly rotating black hole in long GRBs*," Acta Physica Polonica B Proceedings Supplement, **vol. 13, no. 2**, pp. 261-266, 2020.

[41]  A. Janiuk, N. S. Dehsorkh, and D. Król, "*Self-gravitating collapsing star and black hole spin-up in long gamma ray bursts*," Astron. Astrophys., **vol. 677**, p. A19, 2023.

[42]  P. Cronin-Coltsmann, "*The imaging and discovery of M-dwarf debris discs with ALMA*," PhD Thesis, University of Warwick, 2022.

[43]  N. Basu, *Introduction to Fundamental Astronomy*. Educohack Press, 2025.